\documentclass[%
 reprint,
nofootinbib,
 amsmath,amssymb,
 aps,
]{revtex4-2}

\usepackage[english]{babel}
\usepackage{amssymb,amsmath}
\usepackage{xcolor}

\usepackage{footnoterange}
\usepackage{graphicx}
\usepackage{dcolumn}
\usepackage{bm}

\begin{document}


\title{QUANTUM MODEL OF HYDROGEN-LIKE ATOMS IN HILBERT SPACE
\\
\small{BY INTRODUCING THE CREATION AND ANNIHILATION OPERATORS}
} 

\author{Mehdi Miri}
 \affiliation{Department of Physics, Razi University, Kermanshah, Iran.}%
 \email{mehdi.miri.phys@gmail.com}

\date{\today}

\begin{abstract}
The purely algebraic technique associated with the creation and annihilation operators to resolve the radial equation of Hydrogen-like atoms (HLA) for generating the bound energy spectrum and the corresponding wave functions is suitable for many calculations in quantum physics. However, the analytical approach with series is extensively used based on wave mechanics theory in most of quantum textbooks. Indeed, much More complete than the old solution of Schrödinger's time-independent differential equation (TISE), one can simply earn all quantum information of a system by using the operational method. In addition to earlier two models, including the quantum harmonic oscillator and the total angular momentum, it can undoubtedly be a third fundamental model to solve Schrödinger's eigenvalue equation of the HLA systems in Hilbert space similar to the harmonic oscillator. We will illustrate how systematically making an appropriate groundwork to discover the coherent states can lead to providing the energy quantization and normalized radial wave functions attached to the matrix representation without additional assumptions.
\end{abstract}

\maketitle
\section{Introduction}

\indent
The Hydrogen atom is a fundamental problem whose energy eigenvalues and eigenfunctions are at the starting point of many calculations or approximations. The ladder operators are a standard technique that has been widely used to untangle quantum problems. They have the distinct advantage of producing compact results compared to a series solution of Schrödinger equation as it appears to resolve a harmonic oscillator using the ladder operators. In addition, operational algebra is utilized to introduce essential techniques in modern quantum theory and other fields 
\cite{2007AmJPh..75..629M, 2007IJQC..107.1608M}.
Since the eigenvalue problem of the Hydrogen atom related to the radial wave functions is analytically solved for a particle in the coulomb field, we are trying to see using the algebraic solution of the ladder operators in complex coordinates. Most of the research in this context is based on the factorization of the radial Hamiltonian. Schrödinger introduced the factorization of energy eigenequation of one-dimensional harmonic oscillators. Infeld and Hull later extended it, but this method and the concept of ladder operators are used further in the supersymmetry (SUSY). There is a possibility to build various creation and annihilation operators for connecting the nearest neighboring eigenstates \cite{1997PhLA..231....9L}.
The bound states of the Hydrogen atom are of great importance in both classical and quantum mechanics, and new approaches can be used to solve such problems. 

The nonlinear Lie algebra theory can equip the vector ladder operators for three-dimensional isotropic harmonic oscillators and Hydrogen atoms without using the factorization method \cite{1980AmJPh..48..855D, 1999AcPSn...8....8Z}. 
It has long been known that the degeneracy of bound states of non-relativistic Hydrogen atom may be described in terms of an isomorphic group $SO(4)$. The degeneracy of the eigenvalue spectrum is associated with the dimensions of irreducible representations in the symmetry group. Suppose the eigenfunctions belonging to different irreducible representations of a group are always degenerate. In that case, we can say that some symmetries have been ignored, and the system's symmetry must be more extensive than what we have found. There is the degeneracy of the same energy for the HLA systems\footnote{It involves a charged nucleus $+Z e$ with only one electron:\\
$Z=1\rightarrow H,~Z=2\rightarrow He^+,~Z=3\rightarrow Li^{+2},~Z=4\rightarrow Be^{+3},~Z=5\rightarrow B^{+4},~Z=6\rightarrow C^{+5},...$} with different values $l$ but the same value $n$ whatever the symmetry group is more significant than $SO(3)$ \cite{joshi2018elements}. Now, it should mention that total degeneracy would be in the order of $2n^2$ by taking into account the electron spin. 

The potential operator has a spherical symmetry invariant under any symmetry group $SO(3)$. Hamiltonian is invariant under the symmetry group $SO(4)$, which leads to the correct degeneracy of the energy spectrum. Using the larger symmetry group $SO(4)=SO(3)\bigotimes SO(3)$ or Lie group algebra $SU(1,1)$ will cause a closed algebra, but the ladder operators cannot be defined \cite{joshi2018elements, wigner2013group}.

This work discusses the HLA systems as a reference model in a non-relativistic regime and is limited to the bound energy levels without the spin corrections and the perturbation effects. A Hamiltonian operator carries it out with a purely operational technique in Hilbert space similar to the formalism of the quantum harmonic oscillator and the angular momentum.

When P. Dirac first invented the ladder operators to analyze the total angular momentum, at the same time Heisenberg and Schrödinger introduced the quantum harmonic oscillator model and relying on these algebraic methods, W. Pauli derived the Hydrogen atom spectrum in 1926 \cite{1926ZPhy...36..336P}. We can investigate a quantum system despite such a particular approach by combining a pair of the creation and annihilation operators. The operators directly provide the energy eigenvalues and the eigenfunctions of a system without introducing the generative functions and recursive relationships or considering approximate behavior and convergence. On the other hand, Schrödinger discovered a set of Gaussian states for the quantum harmonic oscillator, which do not spread in time and follow the motion of quasi-classical oscillators. In contrast, these states are now called coherent states. He speculated that a similar set of these states for the Hydrogen atom could be constructed. But so far, it has been assumed that such a result can not achieve because, for the harmonic oscillator, the energy eigenvalues are integer space while not for the Hydrogen atom (Nieto and Simmons 1979) \cite{1984JPhA...17L.737G}.

We will illustrate the commutation relations in pairs by a set of Hamiltonian and ladder operators. It is possible to find a closure algebra relation for the HLA systems in which the creation and annihilation operators are defined to take the main steps for constructing coherent states.

\section{Hilbert space and Hamiltonian formalism}
\indent

First and foremost, using analysis of Hamiltonian formalism after rewriting the TISE in a new coordinates system, we determine the creation and annihilation operators. You know, Hilbert space as a state's space is made by finding the commutative operators of the system and tensor product of the corresponding states.  
By accurate modifications and the linear combination of non-commutative operators $r,P_r,$ we try to extract the ladder operators from Hamiltonian operator. These ladder operators will be introduced in the general form $H=k_1 A^\dagger A + k_2$ where $k_1,k_2 \in \mathbb R$.  
In this work, the set of our necessary operators are 
$\left \{H,P_r,L^2,L_i,r,... \right \}$.\\
The commutation relations are given by 
\cite{zettili2009quantum, gasiorowicz2007quantum, sakurai2020modern}:
\begin{equation}
\resizebox{1.0\hsize}
{!}{$\begin{cases}
      \text{$\left[H,L_i\right]=0$}\\
      \text{$[L_i,P^2]=[L_i,r^2]=0$}\\
     \text{$ [L^2,Li]=[L^2,P_r]=[L^2,r]=0$} \\
    \text{$ [L_i,L_j]=i\hbar \varepsilon_{ijk} L_k, ~ [L_i,x_j]=i\hbar \varepsilon_{i j k} x_k,
   ~[L_i,P_j]=i\hbar \varepsilon_{i j k} P_k$}
    \end{cases}$}
    \label{eq:commutations}
\end{equation}
The commutative operators of the quntum system are $\left \{H, L^2, L_z \right \}$ which involve the eigenvalue equations as follow \cite{zettili2009quantum, gasiorowicz2007quantum, sakurai2020modern}:   
\begin{equation}
    \begin{cases}
    ~  \text{$ H \mid n l m_l \rangle = E_n \mid n l m_l           \rangle $},~ n \in \mathbb{N}\\
    ~   \text{$ L_z \mid n l m_l \rangle = 
          \hbar m_l \mid n l m_l \rangle $},~  
     -l \leqslant m_l \leqslant +l \\
     ~  \text{$ L^2 \mid n l m_l \rangle = \hbar^2 l(l+1) \mid n l m_l \rangle$},~ l=0,1,2,...,n-1
    \end{cases}  
    \label{eq:commutations}
\end{equation}

By Hilbert space is made from the joint eigenstate of the operators $H, L^2, L_z$ and specifying three quantum numbers $(n l m_l)$ corresponding to the wave functions $\psi_{nlm_l}$, the eigenstate of the HLA systems can be defined as $\left \vert n l m_l \right \rangle$.
Hamiltonian can be written as a relation in other operators, including the radial momentum operator $P_r,$ the orbital angular momentum $L$, and the radial position operator $r$.  
According to the initial definition of Hamiltonian, we will have:     
\begin{equation}
    H=\frac{P^2}{2\mu} + V(r)
\end{equation}                                  
here, the reduced mass of an electron is 
$\mu = \frac{m_e M}{M+m_e}$.\\ 
The radial momentum operator $P_r$ is given by a Hermitian form \cite{zettili2009quantum}:
\begin{equation}
    P_r=\frac{1}{2}\left[\left(\frac{\mathbf{r}}{r}\right)\cdot \mathbf{P}+ \mathbf{P}\cdot \left(\frac{\mathbf{r}}{r}\right)\right]=-i\hbar\frac{d}{r d r}r
\end{equation}                                 
Using the spherical potential $V(r)= -{Z e^2}/{r}$, Hamiltonian operator is:                               
\begin{equation}
    H_L=\frac{P_r^2}{2\mu}+\frac{L^2}{2\mu r^2}-\frac{Ze^2}{r}
    \label{eq:5}
\end{equation}                                   
The second term in Eq.~\ref{eq:5} is the rotational kinetic energy.
The spatial wave functions consist of the radial functions and spherical harmonics as follows: 
\begin{equation}
    \psi_{n l m_l}\left( \mathbf{r} \right)=R_{n l}\left(r\right)Y_{n l m_l} \left(\theta,\varphi\right)
\end{equation}                                   
If we employ the radial wave functions and the Hamiltonian operator $H_L$, Schrödinger's equation reads:
\begin{align}
    H_{L}R_{n l}= E_{n}R_{n l}
    \end{align}
    \begin{align}
    \left [\frac{d}{d r}(r^2\frac{d}{d r})+\frac{2\mu}{\hbar^2}r^2(E_{n}+\frac{Z e^2}{r})-l(l+1) \right]
    R_{n l}(r)=0 
\end{align}
The completeness and orthogonality conditions of the radial wave functions
$R_{n l}(r)$ are given by \cite{zettili2009quantum, gasiorowicz2007quantum}:
\begin{align}
    \int_0^\infty r^2 {d} r~  {\left|r\right\rangle}{\left\langle r\right |}=1 
    \end{align}
    \begin{align}
    \int_0^\infty r^2 {d} r~R_{\acute{n}l}(r)R_{n l}(r)=\delta_{\acute{n}n}  \label{eq:10}
\end{align}
Let us introduce the dimensionless variables as follows:
\begin{align}
     n=\frac{Z e^2}{\hbar}\sqrt{-\frac{\mu}{2E_n}} 
     \label{eq:11} 
     \end{align}
     \begin{align}
    \rho=\frac{2Z}{n a_\circ} r
     \label{eq:12}
\end{align}
Here, Bohr's radius is $ a_\circ= \frac{\hbar^2}{\mu e^2} \approx 0.53 ~A^\circ$.\\
We need to know that by applying the numerical operator 
$N\equiv\left(-2\frac{\mu a_\circ^2}{Z^2 \hbar^2}H_L \right )^{-1/2}$, the Hamiltonian operator is modified in the new coordinate system by:
 \begin{equation}
     H_N=-\frac{1}{\hbar^2}\rho^2 P_{\rho} ^2-\frac{1}{4}\rho^2+\rho N
      \label{eq:13}
 \end{equation}
The Hermitian numerical operator $N$ can be rewritten as the following relation: 
 \begin{equation}
     N=\frac{1}{\hbar^2}\rho P_{\rho} ^2+\frac{1}{4}\rho+\frac{l(l+1)}{\rho} 
 \end{equation}
Where the radial momentum operator is 
$P_\rho= -i\hbar \frac{d}{\rho d \rho} \rho$. 

What subsequently will be determined is the exact relationship between the continuous basis in the new and old coordinates by the real coefficients $\xi_n,$ meaning that:
\begin{align}
     \left|r \right\rangle&=\xi_n \left|\rho \right\rangle
     \end{align}
     \begin{align}
     \xi_n &=\frac{1}{\sqrt{2n}} \left(\frac{2Z}{n a_\circ} \right )^{3/2}  \label{eq:16}
\end{align}
Considering $R_{n l}(r)= \xi_n \phi _{n l}(\rho)$ by use of Eq.\ref{eq:16}, the normal radial wave functions $R_{n l}(r)$ reads:
\begin{equation}
    R_{n l}(r)=\frac{1}{\sqrt{2n}}\left(\frac{2Z}{n a_\circ} \right )^{3/2}~ \phi_{n l}(\rho)
\end{equation}
Utilization of the eigenfunctions $\phi _{n l}(\rho)$ and 
$H_N \equiv \frac{1}{\hbar^2} L^2$, Schrödinger's equation can be rewritten: 
\begin{align}
    H_N\phi_{n l}=l(l+1)\phi_{n l}   \label{eq:18} 
   \end{align}
   \begin{align}
    \left [\rho^2 \frac{d^2}{d \rho^2}+ 2\rho \frac{d}{d \rho}- \frac{\rho^2}{4}+ n\rho- l(l+1) \right]\phi_{n l}(\rho)=0
\end{align}
The unit operator and the orthonormality condition of $\phi _{n l}(\rho)$ are given by:
\begin{align}
      \int_0^\infty \rho {d} \rho~  {\left|\rho      \right\rangle}{\left\langle \rho \right |}=1 
      \end{align}
      \begin{align}
     \int_0^\infty \rho {d} \rho~ \phi_{\acute{n}l}
    (\rho)\phi_{n l}(\rho)=\delta_{\acute{n}n}
     \label{eq:21}
\end{align}
In a linear vector space, a set of the eigenkets $\left \vert {n l} \right \rangle$ with any constant integer $l$ and $n\ge l+1$ can be considered as an orthonormal and complete basis of Hilbert space:
\begin{equation}
    \left \langle \acute {n} l \vert n l \right \rangle= \delta_{\acute{n}n},~
    \sum_n 
    \left \vert n l \right\rangle 
    \left\langle n l \right \vert
    = 1
    \label{eq:22}
\end{equation}

The eigenfunctions $\phi _{n l}(\rho)$ must be well-behaved, single value, continuous finite and squared integrable. Since the probability of finding an electron near the nucleus is greater than elsewhere, it is zero at infinity and $\phi _{n l}(\rho)$ must contain an exponential element in the form of $e^{-\rho/2}$. One can evidently result that
$\rho \phi _{n l}(\rho)$ will vanish in the boundary conditions of $ \rho= 0, \infty$ \cite{weidner1980elementary}.

\noindent 

\textbf{Definition.}
The equivalence classes $L^2(\Omega, \mu)$ as a set of the vector space $\Omega \equiv \mathbb R^ n$ which are contained the continuous and squared integrable eigenfunctions $\psi_1, \psi_2,$ become a Hilbert space when the inner product is defined by: 
\begin{align}
    \forall \psi_1(\eta),\psi_2(\eta) \in L^2(\Omega,\mu), \eta \in \Omega \equiv \mathbb{R} \Rightarrow
    \nonumber \\
 \left \langle \psi_1 \vert \psi_2 \right \rangle_\mu= \int_\Omega {d} \mu (\eta)~ \psi_1^*(\eta)\psi_2(\eta)~~~~
\end{align}
In the new coordinate system, using the measure theory 
$d\mu(\eta)= \omega(\eta) d \eta $ is an element of volume and will be called the density or weight factor \cite{prugovecki2013quantum}. Opting for the measure $d\mu(\rho)= \rho d \rho $, the real eigenfunctions $\phi_{n l}(\rho)$ belonging to the Hermitian operator $H_N$ can be considered as a normalized complete set of the linear vector space:
\begin{align}
    \left \langle 
 \phi_{n l} \vert {H_N} \vert\phi_{n l} 
 \right \rangle=
\left \langle 
{H_N}\phi_{n l}\vert\phi_{n l} 
\right \rangle  \Rightarrow ~~~~~~~\nonumber\\
\int_0^\infty \rho {d} \rho~ \phi_{n l} H_N
   \phi_{n l}= \int_0^\infty \rho {d}\rho~ 
   (H_N \phi_{n l})^{*} \phi_{n l} 
\end{align}
It should note that the Hermitian operator $\rho P_\rho$ will play an essential role in this work.
The ladder operators will be extracted by Eq.~\ref{eq:13} and the factorization technique as follows:
 \begin{equation}
     H_N= -(\mp \frac{i}{\hbar} \rho P_\rho - \frac{1}{2} \rho + N)(\pm \frac{i}{\hbar} \rho P_\rho - \frac{1}{2} \rho + N)+ N(N\pm 1)
 \end{equation}
Using the conjugate operators $A_{\pm}^\dagger= A_{\mp}$, the Hermitian operator $H_N$ reads:
  \begin{equation}
      H_N= -A_{\pm}^\dagger A_{\pm}+ N(N \pm 1)
        \label{eq:26}
  \end{equation}
The linear operators $A_+, A_-$ would be introduced by: 
   \begin{equation}
   A_{\pm}= \pm \frac{i}{\hbar} \rho P_\rho 
   - \frac{1}{2} \rho + N  \label{eq:27}
   \end{equation}
We will demonstrate that the non-Hermitian operators $A_\pm$ are ladder operators who can voluntarily move levels.

\section{vertical ladder operators}

The extracted operators $A_\pm$ are the same ladder functions, which will produce the creation and annihilation operators.

\subsection{Creation and Annihilation operators}
\indent

Now, we should explain the crucial role of the operators $A_\pm$ and propose desired equations.
The operators $A_\pm$ and $H_N$ would be applied in the N-space:
\begin{equation}
    N{\left |{n l}\right\rangle}= n{\left |{n l}\right\rangle}
     \label{eq:28}
\end{equation}
The eigenvalue equation of $H_N$ according to Eq.~\ref{eq:18} can be rewritten by:                                            
\begin{equation}
    H_N{\left |{n l}\right\rangle}= l(l+1){\left |{n l}\right\rangle}  \label{eq:29}
\end{equation}
By using Eqs.~\ref{eq:26} and \ref{eq:29}, acting the terms of $A_-A_+$ and $A_+A_-$ on a basis $\left \vert {n l} \right \rangle$ is:
\begin{equation}
    A_{-}A_{+}{\left |{n l}\right\rangle}= [n(n+1)-l(l+1)]{\left |{n l}\right\rangle}
      \label{eq:30}
\end{equation}
We can modify Eq.~\ref{eq:26} by necessary conversion $n \to n+1$ as following:
 \begin{equation}
   A_{+}A_{-}{\left |{n+1, l}\right\rangle}= [n(n+1)-l(l+1)]{\left |{n+1, l}\right\rangle}  
  \label{eq:31}
 \end{equation}
Let us multiply the above equation by $A_-$ from the right side:
\begin{equation}
    A_{-}A_{+}A_{-}{\left |{n+1, l}\right\rangle}= [n(n+1)-l(l+1)]~ A_{-}{\left |{n+1, l}\right\rangle}
      \label{eq:32}
\end{equation}
By comparing Eqs.~\ref{eq:30} and \ref{eq:32}, one can be seen that 
$ A_- \left \vert {n+1, l} \right \rangle \sim \left \vert {n l} \right \rangle $.    
By choosing the constant factor as $a_{n+1,l}^ -$ and again conversion $n+1 \to n$ we can say:
\begin{equation}
    A_{-}{\left |{n l}\right\rangle}= a^-_{n l}{\left |{n-1, l}\right\rangle}
\end{equation}
It means that the annihilation operator $A-$ is correctly introduced now.
In the same way, let us multiply Eq.~\ref{eq:30} by 
the creation operator $A_+$ from the right side:
\begin{equation}
    A_{+}A_{-}A_{+}{\left |{n l}\right\rangle}= [n(n+1)-l(l+1)]~ A_{+}{\left |{n l}\right\rangle}
     \label{eq:34}
\end{equation}
By comparing Eqs.~\ref{eq:31} and \ref{eq:34}, one can be seen that $ A_+ \left \vert {n, l} \right \rangle \sim \left \vert {n+1, l} \right \rangle $.    
Hence by the constant factor as $a_{n l}^+$   we write:
\begin{equation}
    A_{+}{\left |{n l}\right\rangle}= a^+_{n l}{\left |{n+1, l}\right\rangle}
\end{equation}

In other words, the creation operator $A_+$ is also introduced perfectly.
The ladder operators $A_{\pm}$ such as a rising and lowering elevator with the eigenvalue $a_{n l}^\pm$ can move on the vertical levels. In other words, since the operators $A_\pm$ are neither Hermitian nor anti-Hermitian, these expectation values should be zero, meaning that 
$\left \langle A_\pm \right \rangle _{n l} =0$. 

\subsection{Eigenvalues of ladder operators}
\indent

Let us find the eigenvalues of ladder operators $A_\pm$ using the orthonormality relation of the system. Therefore, as mentioned more by Dirac notation in the basis $\left \vert {n l} \right \rangle$, it is obvious that:
\begin{equation}
 \overline{A_{\pm}^\dagger A}_{\pm}\equiv
  {\left\langle A_{\pm}^\dagger A_{\pm}\right \rangle}_ {n l}= |a^\pm _{n l}|^2
\end{equation}
The expectation value of the Hermitian operator $A_\pm^\dagger A_\pm$ can be determined using Eqs.~\ref{eq:26} and \ref{eq:29}, that meaning:
\begin{equation}
 \overline{A_{\pm}^\dagger A}_{\pm}=
   n(n\pm 1)-l(l+1)
\end{equation}
Since these expectation values are real, they can be obtained without any restrictive conditions:
\begin{equation}
   a^\pm _{n l}= \left[n(n\pm 1)-l(l+1)\right]^{1/2}
    \label{eq:38}
\end{equation}                                              
Hence, doing straightforward processes, the creation and annihilation operators $A_\pm$ as the linear differential equations are valued.

\section{Discussion}
In the next step, it is thoroughly necessary to acquire 
the energy quantization and wave functions of the HLA systems.

\subsection{Energy quantization}
\indent

We recognize no eigenstate corresponding to the non-integer values $n$. On the other hand, according to the variable definition $\rho$ in Eq.~\ref{eq:12} the negative or zero values $n$ is impossible. Since the eigenvalues of the ladder operators are limited in the condition $\vert a_{n l}^\pm \vert ^2 \ge 0$, it is better to rewrite Eq.~\ref{eq:38} by:
\begin{equation}
   a^\pm _{n l}= \sqrt{(n\mp l)(n\pm l \pm 1)}
\end{equation}
It is assumed $n=n_r+l+1$ where $l=0,1,2,...\in \mathbb Z$ can be called the orbital quantum number. Moreover, the non-negative integer $n_r$ is the radial quantum number.\\
The ground states $\left \vert {l+1,l} \right \rangle$ as an infinite set of kets $\{\left \vert {10} \right \rangle, \left \vert {21} \right \rangle, \left \vert {32} \right \rangle, ... \}$ are the lowest arbitrary floor. When the eigenvalue $a_{l+1,l}^-$ is zero, the sequence will be terminated and the process of acting the operator $A_-$ should be stopped, meaning that $A_- \left \vert {l+1,l} \right \rangle =0$.

The bound energy spectrum $E<0$ by using Eq.~\ref{eq:11} will be extracted by:
\begin{equation}
    E_n= -\frac{1}{2}\mu C^2 \frac{(Z\alpha)^2}{n^2},~n=1,2,3,...\in \mathbb{N}
\end{equation}
The fine-structure constant is $\alpha =\frac{e^2}{\hbar C}=1/137$. \\
We can remind that the derived result entirely agrees with Bohr's theory. 

\subsection{Matrix representation and Radial wave functions}
\indent

The structure of Hilbert space is a highly convenient and powerful tool to provide insight that completes all information offered by the wave functions.

The matrix elements of the operators $\{ H_N,A_\pm,\rho \}$ can be measured in the complete basis $\left \vert {n l} \right \rangle$ of Hilbert space as follows:
\begin{align}
     {\left\langle {\acute{n} l}\right|H_N}
     {\left |{n l}\right\rangle}= 
     l(l+1) \delta_{\acute{n} n} 
     \end{align}
     \begin{align}
        {\left\langle {\acute{n} l}\right|A_{\pm}}
   {\left |{n l}\right\rangle}= [n(n \pm 1)-l(l+1)]^{1/2} \delta_{\acute{n}, n \pm 1} 
     \end{align}
   \begin{align}
     {\left\langle {\acute{n} l}\right|\rho}
    {\left |{n l}\right\rangle}= + 2n\delta_{\acute{n} n}
    -[n(n+ 1)-l(l+1)]^{1/2} \delta_{\acute{n}, n+ 1}
    \nonumber\\
    - [n(n- 1)-l(l+1)]^{1/2} \delta_{\acute{n}, n - 1} 
     ~~~~~~~~~~~
   \end{align}

\begin{figure*}[ht]
\centering
\includegraphics[width= 16cm, height= 7.4cm]
{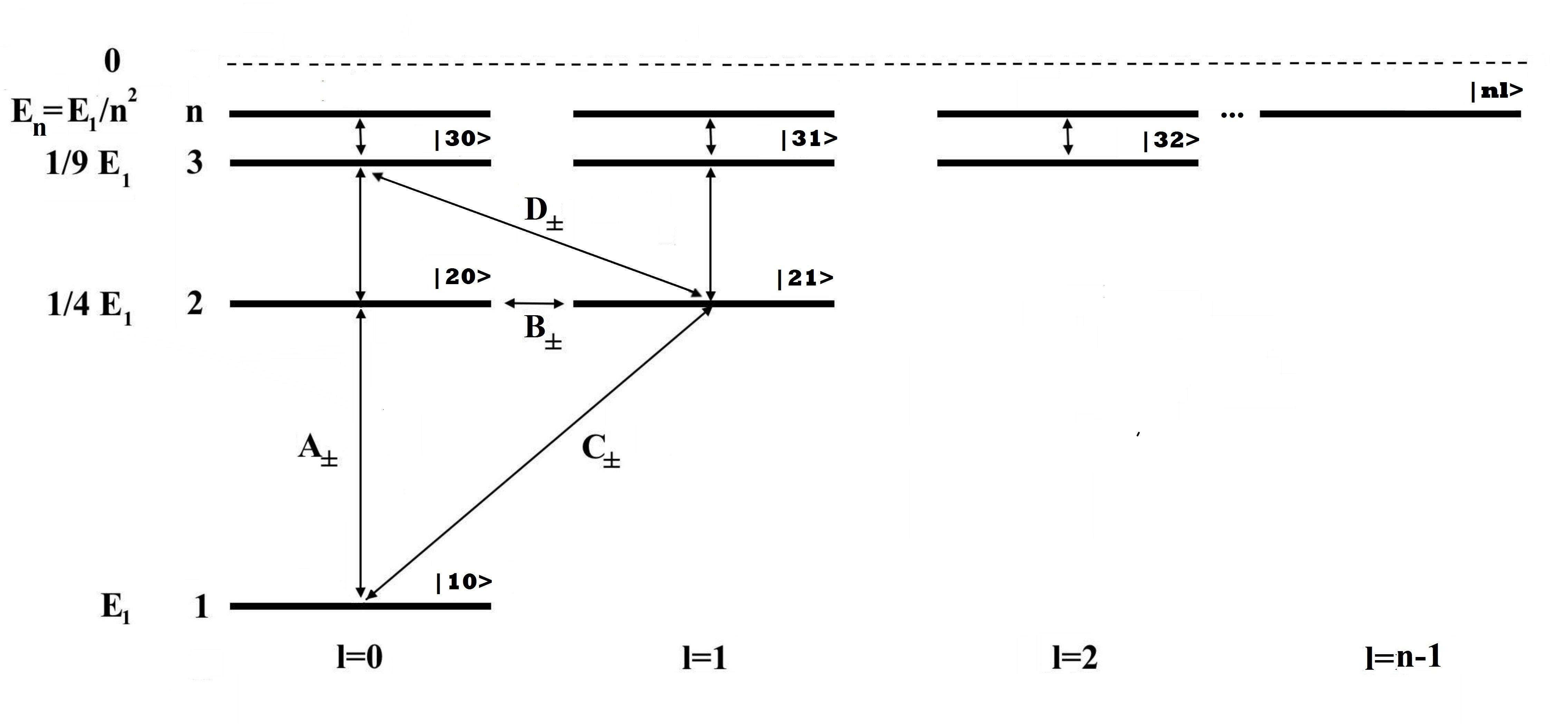}
\caption{The representation of ladder operators $A_\pm, B_\pm, C_\pm, D_\pm$ on the energy levels $\left \vert {n l} \right \rangle$, $E_1 \approx -13.6~\rm ev$.}
\label{fig:levels}
\end{figure*}

Because the operators $N, H_N$ are commutative, they have joint eigenstate and will be represented by infinite diagonal matrices. But the operators $A_\pm$ do not commute with $N$ and are not diagonal in the  N-representation, meaning that $[N,A_\pm]= \pm A_\pm$.   
The eigenvalue equation of $N$ and the relation 
$NA_\pm= A_\pm(N\pm 1)$ can lead to the following result:                                                                                   %
\begin{equation}
    NA_{\pm}{\left |{n l}\right\rangle}= (n \pm 1) A_{\pm}{\left |{n l}\right\rangle}
\end{equation}
you know that $ A_\pm \left \vert {n l} \right \rangle$ are the eigenstates $N$ with the eigenvalues $(n\pm 1)$. 
When $ A_\pm$ act on the state $\left \vert {n l} \right \rangle$, they increase or decrease by one unit, and a new state as 
$\left \vert {n\pm 1, l} \right \rangle$ will be generated, meaning that $ A_\pm \left \vert {n l} \right \rangle \sim 
\left \vert {n\pm 1, l} \right \rangle $.
It is worth noting that since the commutation relation $A_\pm$ by Eq.~\ref{eq:26} is as $[A_-, A_+]=2N$, we have:
  \begin{equation}
    N= \frac{1}{2} (A_{-}A_{+} -A_{+}A_{-}  )
    \label{eq:45}
  \end{equation}                                          %
Specifically, the following results related to the vertical ladder operators $A_\pm$ can be obtained:
\begin{equation}
    A_{\pm}{\left |{n l}\right\rangle}= [n(n \pm 1)-l(l+1)]^{1/2}{\left |{n \pm 1, l}\right\rangle}
\end{equation}
One can simply achieve any normal eigenfunctions $\phi_{n l} (\rho)$ by solving a first-order linear differential equation as follows:
\begin{align}
     A_{\pm}\phi _{n l}(\rho)= (\pm \rho \frac{d}{d \rho}- \frac{\rho}{2}+ n \pm 1)\phi _{n l}(\rho) 
     ~~~\nonumber \\
     =\sqrt{(n\mp l)(n \pm l \pm 1)}\phi _{n \pm 1, l}(\rho) 
    \label{eq:47}
\end{align}
We will find $\phi_{n l} (\rho)$ by acting $A_\pm$ on any energy level $\left \vert {n l} \right \rangle$ as shown in Fig.~\ref{fig:levels}. 
Individually by increasing or decreasing one just unit of various quantum numbers $n, l, n_r$, the desired ladder operators can be defined.

The operators $A_\pm$ connect the nearest neighboring eigenstates with the same angular momentum but different energy, meaning that the orbital quantum number $l$ will be constant. It is possible to make a pair of the horizontal ladder operators $B_\pm$ as right-handed or left-handed operators which connect the nearest neighboring eigenstates with the same energy but a different angular momentum meaning that the main quantum number $n$ should be constant. On the other hand, the operators $C_\pm$ connect the nearest neighboring eigenstates with the same radial quantum number $n_r$ and the operators $D_\pm$ connect the nearest neighboring eigenstates with the same $n+l$ automatically \cite{1997PhLA..231....9L, 1999PhLA..259..212X, de1991operator}.

Now, we will detect the radial wave functions $R_{n l}(r)$ related to the HLA as a general formula.
The ground states $\phi_{l+1, l} (\rho)$ by substituting $n=l+1$ would be extracted of the following equation:
\begin{equation}
    A_{-}\phi _{l+1, l}(\rho)= (- \rho \frac{d}{d \rho}- \frac{\rho}{2}+  l)\phi _{l+1, l}(\rho)= 0 
\end{equation}
Based on the normalization relation in Eq.~\ref{eq:21}, the general solution $\phi_{l+1, l} (\rho)$ is given by:
\begin{equation}
    \phi _{l+1, l}(\rho)= \frac
    {1}
    {\sqrt{(2l+1)!} }
    \rho^ l e^{-\rho/2}
\end{equation}

Here, the excited states $\left \vert {n l} \right \rangle$ will be consecutively ensued from acting the creation operator $A_+$ on the ground states $\left \vert {l+1, l} \right \rangle$ in order of 
$(n-l-1)$ as a recurrence relation with any constant integer number $l$, in which will be used the eigenvalues $a_{k l}^+$ where 
$k=l+1,l+2,...,n-2,n-1$ as follows:
\begin{align}
     {\left |{n l}\right\rangle}=
     \frac
     {1}{a^+_{n-1, l}
     a^+_{n-2, l}~ \cdots
     a^+_{l+1, l}}
     A_+ A_+~ \cdots  A_+
    {\left |{l+1, l}\right\rangle} 
    \nonumber \\
  =
  \prod_{k=l+1}^{n-1}
  \frac{1}
  {\sqrt{(k-l)(k+l+1)}}
  A^{n-l-1}_+
    {\left |{l+1, l}
    \right\rangle}
      ~~~~
     \label{eq:50}
\end{align}
   The solution of the above product converge series reads:
\begin{equation}
     {\left |{n l}\right\rangle}= 
     \sqrt{\frac{
     (2l+1)!}{(n+l)!(n-l-1)!}}
     ~  A^{n-l-1}_+ {\left |{l+1, l}\right\rangle} \in L^2(\mathbb{R})
     \label{eq:51}
\end{equation}

In continuation of the discussion, the ahead procedure is an intrinsically interesting result because we don’t need to solve the complicated Laguerre equation. The relationship between the eigenfunctions $\phi_{n l} (\rho)$ and the associated Laguerre's polynomials meaning that $L_{n+l}^ {2l+1} (\rho)$ is 
\cite{pauling2012introduction}:
\begin{equation}
    \phi _{n l}(\rho)= -C_{n l} \rho^l e^{-\rho/2} L^{2l+1}_{n+l}(\rho) 
\end{equation}
The Laguerre differential equation is not self-adjoint. As an important result of the Sturm-Liouville theorem and considering appropriate weight factor, the orthonormality condition of the associated Laguerre polynomials is \cite{arfken2013mathematical, boas2006mathematical}:
\begin{equation}
   \int_0^\infty \rho d \rho~  \rho^{2l}e^{-\rho} L^{2l+1}_{\acute{n}+l}(\rho)L^{2l+1}_{n+l}(\rho)= \frac{[(n+l)!]^3}{(n-l-1)!}\delta_{\acute{n}n}
\end{equation}
Using the normalization condition of $\phi_{n l} (\rho)$, the constant $C_{n l}$ is determined by:
\begin{equation}
     \phi _{n l}(\rho)=  
      -{\left [ \frac{(n-l-1)!}{[(n+l)!]^3} \right ]^{\frac{1}{2}}}
     \rho^l
     e^{-\rho/2}  L^{2l+1}_{n+l}(\rho)
\end{equation}
Because $R_{n l} (r)= \frac{1}{\sqrt{2n}} \left ( \frac{2Z}{n a\circ} \right)^{3/2} \phi_{n l} (\rho) $ and $\rho= \frac{2Z}{n a_\circ}r $, 
the radial wave functions $R_{n l}(r)$ can be written in the following form:
\begin{align}
     R _{n l}(r)= 
     - {\left(\frac{2Z}{n a_\circ}\right)^{\frac{3}{2}}} 
     {\left [ \frac{(n-l-1)!}{2n[(n+l)!]^3} \right ]^{\frac{1}{2}}} {\left(\frac{2Z}{n a_\circ}r \right)^ l}
      \nonumber \\
    \times
     L^{2l+1}_{n+l}{\left(\frac{2Z}{n a_\circ}r \right)} e^{\frac{-Z}{n a_\circ}r}
     ~~~~~~~~~~~~~~~~~~~~
\end{align}
The distribution curve of probability density function
$p(r)= r^2 \vert R_{n l}(r) \vert ^2 $ has $(n-l)$ bulges. 
The expectation value $r^k$ which is a special parameter would be achieved as follows \cite{zettili2009quantum, gasiorowicz2007quantum}:
\begin{equation}
     {\left\langle r^k \right \rangle}_ {n l}= 
     \int_0^\infty d r~ r^{2+k} \vert R_{n l}(r) \vert ^2
\end{equation}
By the properties of Laguerre polynomials, we show 
\cite{zettili2009quantum}:
\begin{align}
    {\left\langle r \right \rangle}_ {n l}= 
     \frac{a_\circ}{2Z}[3n^2 - l(l+1)] 
     \end{align}
     \begin{align}
    {\left\langle r^{-1} \right \rangle}_ {n l}= 
     \frac{Z}{a_\circ n^2}
\end{align}
According to Bohr's theory, the maximum Probability of electron presence is in the ground state of $R_{n, n-1}(r)$, where we see the highest bulge. It is $\overline{r} \sim r_n= a_\circ n^2 $ as the quantized radius of circular orbits for the Hydrogen atom.\\ 
The general eigenstates of the HLA systems, which have a spatial and spin part, are given by:
 \begin{equation}
     \left \vert n l m_l m_s \right \rangle= 
     \left \vert n l m_l \right \rangle \bigotimes
     \left \vert s;~ m_s \right \rangle 
 \end{equation}
The spatial functions $\psi_{n l m_l}$  in the spherical coordinate system consist of the radial wave functions 
$R_{n 1}(r)$ and angular dependence $Y_{l m_l}(\theta, \varphi)$, which will result in:
\begin{align}
     \psi_ {n l m_l} (r, \theta, \varphi)=
    - {\left(\frac{2Z}{n a_\circ}\right)^{3/2}} 
     {\left [ \frac{(n-l-1)!}{2n[(n+l)!]^3} \right ]^{1/2}} {\left(\frac{2Z}{n a_\circ}r \right)^ l}
    \nonumber \\
    \times
     L^{2l+1}_{n+l}{\left(\frac{2Z}{n a_\circ}r \right)} e^{-\frac{Z}{n a_\circ}r}~ Y_ {l m_l} (\theta, \varphi)
     ~~~~~~~~
      \label{eq:60}
\end{align}
Subsequently, the perfect wave eigenfunctions are:
 \begin{equation}
    \psi_ {n l m_l m_s} (\overrightarrow{r})=
    \psi_ {n l m_l} (r, \theta, \varphi)
     \left \vert 1/2;~ \pm 1/2 \right \rangle
 \end{equation}
These wave functions 
can be checked in most quantum physics textbooks and other available articles to study more thoroughly.

\subsection{Coherent states}
\indent

We shall go on to another set of basis states known as the coherent states or Bargmann's representations, which have far more physical significance than the technical aspects we have already studied. It is only recently that such a concept can be generalized to quantum field theory in depth. The importance of this matter is expanded simultaneously with the fast development of technologies in lasers, photonics science, and quantum optics in recent decades. Not only does the coherent state play a crucial role in new technology, but it also comes with a multitude number of benefits in the future.
The coherent states $\left \vert z \right \rangle$ as either minimum uncertainty states or a set of all eigenvectors of the annihilation operator satisfy the following equation:
\begin{equation}
    A_- \left \vert z \right \rangle =
    z  \left \vert z \right \rangle
\end{equation}
They are labeled with the complex numbers 
$z=\vert z \vert e^{i K} $ as general solutions of the eigenvector equation because the operator $A_-$ is not a Hermitian operator \cite{1963PhRv..131.2766G, baulieu2017classical}. 
By expanding the normalized eigenstates $\left \vert z \right \rangle$, meaning that $\left \langle z \right \vert z \rangle =1$ on the standard orthogonal basis $\left \vert {n l} \right \rangle$ according to the relation $\left \vert z \right \rangle =
   \sum_{n} c_n
   \left \vert {n l} \right \rangle$ 
and also using Eqs. 22 and 51, we can solve all possible coherent states for any constant integer
$l,~n \ge l+1$ as follows:
\begin{equation}
   \left \vert z \right \rangle =
   {M_l} \sum_{n}
   \frac{z^{n-l-1}}{\sqrt{(n+l)!(n-l-1)!}} 
   \left \vert {n l} \right \rangle
\end{equation}
The modified Bessel’s equation of the first kind differs from Bessel ODE but only a small change $x \rightarrow ix $ will be caused to construct the Hyperbolic Bessel functions by an imaginary argument as 
$I_l(x)\equiv e^{-i \pi l /2} J_l(ix) = \sum_{k=0}^ \infty 
\frac{\left(\frac{x}{2}\right)^{2k+l}}{k!(k+l)!} $ for integer numbers $l,k \ge 0$ and all real values $x \equiv 2 \vert z \vert \in \mathbb{R}$. The modified functions $I_l(x)$ are not oscillatory but have exponential behavior in character (rather than trigonometric). Fortunately, the knowledge that we have developed regarding Bessel ODE can be well used for modified Bessel functions. Here, i t is necessary to note that real recurrence coefficients ${M_l}$ are given by \cite{arfken2013mathematical, korenev2019bessel}:
    \begin{align}
      {M_l}(x)  \equiv 
      \left (  \frac{x}{2} \right )^ {l+1/2} 
      \left [I_{2l+1}(x) \right ] ^ {-1/2}
      \nonumber \\
      =\left [\sum_{k=0}^ \infty  
       \frac{\left (\frac{x}{2} \right )^{2k}}{k!(k+2l+1)!}
      \right ]^{-1/2}
     ~,k=0,1,2,... \in
     \mathbb{Z}
      \label{eq:64}
 \end{align} 
As we all know, the power series expansion of $I_l(x),~ M_l(x)$ is similar to $J_l(x)$ but without the alternating factor $(-1)^m$.
\begin{figure*}[ht]
\centering
\includegraphics
[width=0.8\hsize,clip]{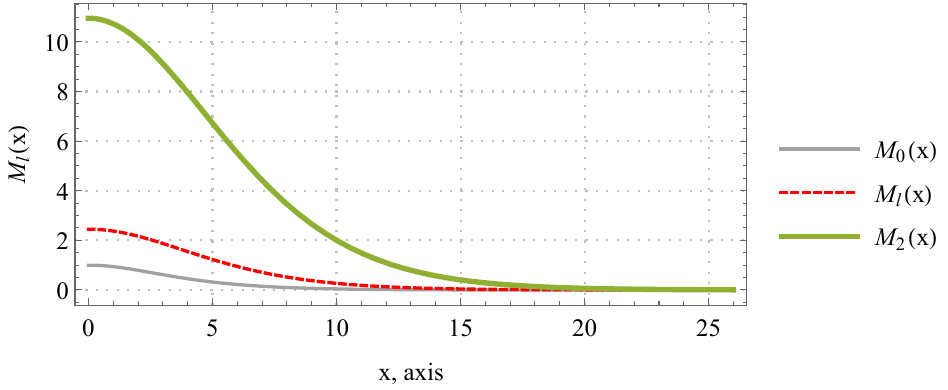}
\caption{
The Bessel's function of the first kind can be defined by contour's integral as 
$J_l(x)=\frac{1}{\pi}\int_0^\pi{d}t~ \cos(x\sin t-l t) $ which is valid even for complex arguments meaning that an important special case with a purely imaginary argument $(ix)$. Now, the solutions of Bessel's equation are called the modified functions (or occasionally the hyperbolic Bessel's functions). \\
The functions $J_l(ix)$ will be real for all even integers $l$ and take imaginary values for odd integers $l$. To get rid of this simple issue, one introduces into consideration the modified Bessel's functions $I_l(x)= i^{-l}J_l(ix)$ which are real for any $l$ and also closely related to $J_l(x)$. These entire functions for large values $x$ and integer numbers $l$ can behave similarly to an exponential function of a real positive argument. It is crucial to be noted that unlike the ordinary Bessel's functions $J_l(x)$ which are oscillating, the functions $I_l(x)$ and $M_l(x)$ are exponentially growing and decaying graphs respectively. we know that the functions $I_l(x)$ will go to zero at $x=0$ for $l >{0}$ and are finite at $x=0$ for $l=0$ . The convergent series of $M_l(x)$ is primarily involved in a descending sequence as $a_k=\frac{1}{4^k (k+2l+1)!~k! }$. Since $ \lim_{k \to +\infty}\vert \frac{a_{k+1}}{a_k}\vert=0 $, we can find that the rate of convergence with the radius $R=+\infty$ at the infinite interval $-\infty <x< +\infty$ converges to zero super-linearly and very quickly. Here, modified Bessel's functions $ M_l (x) $ have been finite at $x> 0$ for all $l$ despite a singularity of Gaussian type in very far points but absolutely on the boundary conditions of $ x= 0 $ for any odd integers $l$ are as a one-sided limit.}
\label{fig:bessel}
\end{figure*}

 What truly matters for describing quantum physics is the existence of the completeness relation \cite{baulieu2017classical}:
\begin{equation}
    \sum_z  \left \vert z \right \rangle 
     \left \langle z \right \vert \equiv
     \int d z~ \left \vert z \right \rangle 
     \left \langle z \right \vert =1
\end{equation}
The coherent states consist of an overcomplete basis of Hilbert space as can be seen that these elements are never orthogonal, meaning that $\left \langle \acute{z} \right \vert z \rangle \ne{0}$ \cite{baulieu2017classical}.

We can rewrite a more convenient expression using Eqs.~\ref{eq:51}, and 63 as follows:
\begin{equation}
    \left \vert z \right \rangle =
    D_l(z) \left \vert l+1, l \right \rangle 
\end{equation}
The displacement operator $D_l(z)$ is:
\begin{equation}
    D_l(z)=
  {M_l}{\sqrt{(2l+1)!}}
    \sum_{K=0}^ \infty \frac{{z^k}}{k!(k+2l+1)!} A_{+}^K 
\end{equation}
By acting $D_l(z)$, all normalized coherent states $\left \vert z \right \rangle$ out of the stationary ground states will be created. 
It is more interesting that (with the exception $\left \vert z \right \rangle$ being the vacuum) the coherent states can be projected on any given basis $\left \vert {n l} \right \rangle$ by:
\begin{equation}
    \left \langle n l \right \vert z \rangle =
    \frac{{M_l}z^{n-l-1}}
    {\sqrt{(n+l)!(n-l-1)!}}
\end{equation}
It contains all possible states of the Fock basis, which are distributed with a Poisson probability distribution.

The probability amplitude of getting the ground state $\left \vert {10} \right \rangle$ in any coherent state $\left \vert z \right \rangle$ is:
\begin{equation}
\left \langle 1 0 \right \vert z \rangle = {M_0}
    \end{equation}
On the other hand, the occupation number for the eigenvalues of the Hamiltonian is as random as it can be in any given coherent state through a Poisson distribution \cite{baulieu2017classical}.

Here, the average value of the occupation number $N$ reads:
\begin{equation}
    \overline{N}=
      \vert {M_l} \vert^2 
      \sum_{k=0}^ \infty  
      \frac{(k+l+1)}{k!(k+2l+1)!}
      \vert z \vert^{2k}
      \propto \vert z \vert^2
\end{equation}
Obviously (no matter how small or large), whatever the value of $\vert z \vert$ is, the coherent states $\left \vert z \right \rangle$ constitute a maximally random distribution based on an orthogonal basis, which may predict the important physical attributes.

The following wave functions, normalized to one coherent state, can be computed by:
  \begin{align}
    R_{nl}(z,r)= 
    {M_l} \sum_{n}
    \frac{z^{n-l-1}}
    {\sqrt{(n+l)!(n-l-1)!}}
    R_{n l}(r) 
\end{align}

The above-obtained results are explicitly different from the table of coherent states given by approaches such as the group theory. The corresponding coherent states imply that such as aspect can be developed. The annihilation operator will investigate the quantum statistical behavior of the coherent optical fields. 
We will limit our discussion to a determination of such as initial instructions, while elsewhere, we shall give a more detailed account, including an attempt to obtain more effects. It is more crucial to establish that the coherent state is likewise the smallest deterministic state. Moreover, we will notice shortly that any coherent state is a specific kind of quantum state whose dynamics closely try to simulate the behavior of a classical system.

\subsection{Uncertainty principle}
\indent

It is necessary to introduce the uncertainty principle after displaying the coherent states. Heisenberg discovered the uncertainty principle in 1926. He observed that every pair of non-commuting physical properties may result in an uncertainty relationship. Such an assumption led to the development of quantum mechanics theory. 
We have a system with appropriately normalized wave functions and ladder operators $A_\pm$ that:
 \begin{equation}
     A_{\pm}= \frac{1}{\sqrt{2}} (q \mp i p)
     \label{eq:72}
 \end{equation}
The uncertainty in two Hermitian dimensionless operators $q,p$ will be defined by taking into consideration $\sigma_q \equiv \Delta q$ and
 $\sigma_p \equiv \Delta p $ as follows:
\begin{equation}
    \sigma_q \sigma_p \ge \frac{1}{2} 
    \vert \left \langle [q,p] \right \rangle \vert
\end{equation}
This matter is called Heisenberg's uncertainty principle.

Using Eqs.~\ref{eq:45} and \ref{eq:72}, since $[A_-,A_+]= \frac{1}{i}[q,p]= 2N$,    
and the operators $q,p$ are symmetric, the uncertainty relationship can be applied to them:                          
\begin{equation}
   \sigma_q  \sigma_p   
    \ge \overline{N}
\end{equation}
Based on the expectation value of occupation number $N$, the indeterminacy is related to complex variables $z$ and the quantum numbers $n, l$.  
Our primary goal is to find the wave functions known as the least indeterminacy states that satisfy Heisenberg's uncertainty principle with equality. 

By considering Eqs. 45, 62, and 72, the uncertainty of the symmetric operators $q,p$ in the state $ \left \vert z \right \rangle$ reads:
\begin{align}
    \left (\Delta q \right)^2_z \equiv
   { \left \langle q^2 \right \rangle }_z - 
     {\left \langle q \right \rangle}_z^2 = \overline{N}  
     \label{eq: 75}
     \end{align}
     \begin{align}
     \left (\Delta p \right)^2_z \equiv
   {\left \langle p^2 \right \rangle}_z - 
     {\left \langle p \right \rangle}_z^2 = \overline{N}
      \label{eq: 76}
\end{align}
We get the uncertainty relationship with equality by using Eqs.~\ref{eq: 75} and \ref{eq: 76}, which follows:
  \begin{equation}
       \left (\sigma_q \right)_z
        \left (\sigma_p \right)_z = \overline{N}
  \end{equation}
The equal sign implies in the preceding formula that the coherent states
$\left \vert z \right \rangle$ are the least indeterminacy states and 
$R_{nl}(z,r)$ are the minimum uncertainty wave functions.
We can also apply the uncertainty principle on the basis
$\left \vert {n l} \right \rangle$ by Eqs.28 and 74 as follows: 
  \begin{equation}
       \left (\sigma_q \right)_{n l}
        \left (\sigma_p \right)_{n l} \ge n
         \label{eq: 78}
  \end{equation}
Besides, in a similar way by computing all expectation values, the uncertainty of the symmetric operators $q,p$ in the basis  state $\left \vert {n l} \right \rangle$ is:
  \begin{equation}
       \left (\sigma_q \right)_{n l}
        \left (\sigma_p \right)_{n l} = 
        n^2 - l(l+1),~ n \ge {l+1}
  \end{equation}
Hence, the inequality of Eq.~\ref{eq: 78} will be easily extracted as a result of assumption $n^2 -l(l+1) \ge n$.  
The important point that would be cause all wave functions 
$\left \vert {n l} \right \rangle ,~ n \ne l+1$ to have no the minimal indeterminacy is that $A_- \left \vert {n l} \right \rangle \ne 0$.
We know that $\left \vert {n l} \right \rangle$ is not an eigenstate of the operator $A_-$ and it has a non-minimum uncertainty since the uncertainty grows with all values $n \ne l+1$. 

Where the wave functions are orthonormal, the equation
$A_- \left \vert {l+1, l} \right \rangle =0 ,~ n=l+1$ can lead to:
  \begin{equation}
       \left (\sigma_q \right)_{l+1, l}
        \left (\sigma_p \right)_{l+1, l} = l+1
         \label{eq: 79}
  \end{equation}
The comparison of Eqs.~\ref{eq: 78} and \ref{eq: 79} plainly expresses that the ground state
$\left \vert {l+1, l} \right \rangle$ is a minimum uncertainty wave function for any constant value $l$. We can also expect other minimum uncertainty function $ \left \vert z \right \rangle$ due to satisfying the relationship $ A_- \left \vert z \right \rangle = z \left \vert z \right \rangle$.   

\section{Conclusion}
The energy quantization and wave functions have been extracted for the HLA quantum systems by the creation and annihilation operators in Hilbert space without any arbitrary assumptions. The radius and transition relations literally can obtain not using the analytical solution of Schrödinger's equation. It is necessary to consider a few corrections in a much more realistic study of Hydrogen atom, such as one of the most important cases is the relativistic motion of electrons which has caused the emergence of terms in the kinetic energy and We should solve it by the perturbation theory. Incidentally, another correction from a large order of reduced mass will be related to interaction of the nucleus electromagnetic field with the electron spin. We hope 
with an emphasis on research with lasting and profound impact to see more empirical applications of the algebraic method by making an appropriate groundwork to discover the coherent states of Hydrogen atom in the quantum field theory (QFT) and elsewhere. Moreover, it is a brilliant way to produce new technology.

\acknowledgements
After defending my MSc Thesis in the Physics Department of Razi University on July 27, 2018, I stubbornly stuck to this work. 
I would like to admire Prof. M. V. Takook as a supervisor for offering helpful guidance to better progress when I started studying on such an exhausting investigation that lasted about four years.

\appendix
\begin{widetext}
\section*{Appendix A:~Matrix representations of introduced operators in the N-space}

The matrix representation $N,H_N$ for the column $l=0$ in the basis $\left \vert n l \right \rangle $ is:
\begin{equation}
N=
  \begin{pmatrix}
  1 & 0 & 0 & \cdots \\  0 & 2 & 0 & \cdots \\ 0 & 0 & 3 & \cdots
  \\ \vdots & \vdots & \vdots & \ddots
  \end{pmatrix},~~
H_N=
  \begin{pmatrix}
{\left\langle 1 0\right|H_N}{\left | 1 0\right\rangle} & 
{\left\langle 1 0\right|H_N}{\left | 2 0\right\rangle} & 
{\left\langle 1 0\right|H_N}{\left | 3 0\right\rangle} & \cdots
\\{\left\langle 2 0\right|H_N}{\left | 1 0\right\rangle} &
{\left\langle 2 0\right|H_N}{\left |2 0\right\rangle} &
{\left\langle 2 0\right|H_N}{\left | 3 0\right\rangle}& \cdots
\\{\left\langle 3 0\right|H_N}{\left | 1 0\right\rangle} &
{\left\langle 3 0\right|H_N}{\left | 2 0\right\rangle} &
{\left\langle 3 0\right|H_N}{\left | 3 0\right\rangle}& \cdots
 \\ \vdots & \vdots & \vdots & \ddots
  \end{pmatrix}
  =
  \begin{pmatrix}
  0 & 0 & 0 & \cdots \\  0 & 0 & 0 & \cdots \\ 0 & 0 & 0 & \cdots
  \\ \vdots & \vdots & \vdots & \ddots
  \end{pmatrix}  
  \end{equation}
The matrix representation of the vertical ladder operators $A_\pm$  is:
\begin{equation}
A_+=
  \begin{pmatrix}
  0 & 0 & 0 & 0 & \cdots \\  \sqrt{2} & 0 & 0 & 0 & \cdots 
  \\ 0 & \sqrt{6} & 0 & 0 & \cdots
  \\ 0 & 0 &  2\sqrt{3} & 0 & \cdots
  \\ \vdots & \vdots & \vdots & \vdots & \ddots
  \end{pmatrix},~~
A_-=
  \begin{pmatrix}
  0 & \sqrt{2} & 0 & 0 & \cdots \\  0 & 0 & \sqrt{6} & 0 & \cdots 
  \\ 0 & 0 & 0 & 2\sqrt{3} & \cdots
  \\ 0 & 0 & 0 & 0 & \cdots
  \\ \vdots & \vdots & \vdots & \vdots & \ddots
  \end{pmatrix}
  \end{equation}
The joint eigenvectors of $N, H_N$ can be obtained according to Eq.~\ref{eq:28} as follows:
\begin{equation}
N \left \vert {n l} \right \rangle =
  \begin{pmatrix}
  1 & 0 & 0 & \cdots \\  0 & 2 & 0 & \cdots \\ 0 & 0 & 3 & \cdots
  \\ \vdots & \vdots & \vdots & \ddots
  \end{pmatrix}
  \begin{pmatrix}
  a \\  b \\ c \\ \vdots 
  \end{pmatrix}
  = n
  \begin{pmatrix}
 a \\  b \\ c \\ \vdots 
  \end{pmatrix}
  \Rightarrow \begin{cases}
    ~a=n a \\ ~2b=n b \\ ~3c=n c \\ ~\vdots 
  \end{cases}
\end{equation}
The normalized solutions of these equations for $n=1,2,3,...\in \mathbb N$  are respectively given by:
 \begin{equation}
     \left \vert 1 0 \right \rangle =
     \begin{pmatrix}
      1 \\  0 \\ 0 \\ \vdots 
     \end{pmatrix},~
     \left \vert 2 0 \right \rangle =
     \begin{pmatrix}
      0 \\  1 \\ 0 \\ \vdots 
     \end{pmatrix},~
     \left \vert 3 0 \right \rangle =
     \begin{pmatrix}
      0 \\  0 \\ 1 \\ \vdots 
     \end{pmatrix}, ...
 \end{equation}
We can verify that these vectors are orthonormal. Hence according to Eq.~\ref{eq:22}, it is easy to see that:
\begin{equation}
    \left \langle \acute{n}, 0 \vert n, 0 \right \rangle =
    \delta_{\acute{n}n}
\end{equation}
We can also verify that they are complete. Hence according to Eq.~\ref{eq:22}, we have:

\begin{align}
     \sum_{n=1}^N  
    \left \vert n,0 \right \rangle
    \left \langle n,0 \right \vert = 
        \begin{pmatrix}
        1 \\  0 \\ 0 \\ \vdots 
        \end{pmatrix}
        \begin{pmatrix}
        1 & 0 & 0 & \cdots
        \end{pmatrix} +
        \begin{pmatrix}
        0 \\  1 \\ 0 \\ \vdots 
        \end{pmatrix}
        \begin{pmatrix}
         0 & 1 & 0 & \cdots
        \end{pmatrix} +
        \begin{pmatrix}
        0 \\  0 \\ 1 \\ \vdots 
        \end{pmatrix}
        \begin{pmatrix}
        0 & 0 & 1 & \cdots
        \end{pmatrix} + ... =
        \begin{pmatrix}
        1 & 0 & 0 & \cdots \\  0 & 1 & 0 & \cdots 
       \\ 0 & 0 & 1 & \cdots
        \\ \vdots & \vdots & \vdots & \ddots
        \end{pmatrix} \Rightarrow  \nonumber
        \end{align}
        \begin{align}
     \sum_{n=1}^N  
    \left \vert n,0 \right \rangle
    \left \langle n,0 \right \vert = 1
\end{align}
The position and radial momentum operators $\rho, P_\rho$   in terms of $A_\pm$  using Eq.~\ref{eq:27} are given by:
\begin{align}
     \rho= 2N-(A_- + A_+)  
     \end{align}
     \begin{align}
     P_\rho= \frac{i\hbar}{2} \rho^{-1} (A_- - A_+)
\end{align}
The matrix representation of these operators is:            
\begin{equation}
 P_\rho= \frac{i\hbar}{2}
    \begin{pmatrix} 
     -3/5 & 2\sqrt{2}/5 & 2\sqrt{3}/5 & \cdots  
  \\  -3\sqrt{2}/5 & 1/5 & 2\sqrt{6}/5 & \cdots 
  \\ -4\sqrt{3}/15 & -4\sqrt{6}/15 & 1/5 & \cdots
  \\ \vdots & \vdots  & \vdots & \ddots  
    \end{pmatrix},~~
\rho=
    \begin{pmatrix}
     2 & -\sqrt{2} & 0 & 0 & \cdots 
  \\  -\sqrt{2} & 4 & -\sqrt{6} & 0 & \cdots 
  \\ 0 & -\sqrt{6} & 6 & -2\sqrt{3} & \cdots
  \\ 0 & 0 &  -2\sqrt{3} & 8 & \cdots
  \\ \vdots & \vdots & \vdots & \vdots & \ddots 
    \end{pmatrix}
\end{equation}
It is important to be noted that the matrix representation of the operator
$\rho P_\rho$  is:
\begin{equation}
    \rho P_\rho= \frac{i\hbar}{2}
    \begin{pmatrix}
     0 & \sqrt{2} & 0 & 0 & \cdots 
  \\  -\sqrt{2} & 0 & \sqrt{6} & 0 & \cdots 
  \\ 0 & -\sqrt{6} & 0 & 2\sqrt{3} & \cdots
  \\ 0 & 0 &  -2\sqrt{3} & 0 & \cdots
  \\ \vdots & \vdots & \vdots & \vdots & \ddots 
    \end{pmatrix}
\end{equation}
Since the expectation values of $\left \{A_\pm^2,A_\pm,\rho P_\rho \right \}$  in the basis $\left \vert n l \right \rangle $  are zero and also $\left \langle \rho \right \rangle _ {n l}= 2n $ then: 
\begin{align}
    {\left \langle \rho^2 \right \rangle}_{n l}=
    6n^2-2l(l+1) 
    \end{align}
    \begin{align}
     {\left \langle \rho^2 P_\rho^2 \right \rangle}_{n l}=
     \frac{\hbar^2}{2}
    \left [n^2 - l(l+1) \right]
\end{align}
The expectation value $H_N$  in the basis $\left \vert n l \right \rangle $  based on Eq.~\ref{eq:29} reads:
\begin{equation}
    \overline {H}_N= l(l+1)
\end{equation}
We can calculate 
$ {\left \langle (\Delta A)^2 \right \rangle}
{\left \langle (\Delta B)^2 \right \rangle} 
\ge \frac{1}{4} \vert {\left \langle [A,B]\right \rangle} \vert^2
$  according to $ [\rho, P_\rho]=i\hbar $, hence:
\begin{equation}
    \Delta \rho \Delta P_\rho \ge \frac{\hbar}{2}
\end{equation}
What equally matters based on Heisenberg's principle for quantum physics is minimizing the uncertainty distributed in both symmetric operators.\\
The matrix representation $N,H_N$ for the column $l=1$ in the basis $\left \vert n l \right \rangle $ is:
\begin{equation}
N=
  \begin{pmatrix}
  2 & 0 & 0 & \cdots \\  0 & 3 & 0 & \cdots \\ 0 & 0 & 4 & \cdots
  \\ \vdots & \vdots & \vdots & \ddots
  \end{pmatrix},~
H_N=
  \begin{pmatrix}
{\left\langle 2 1\right|H_N}{\left | 2 1\right\rangle} & 
{\left\langle 2 1\right|H_N}{\left | 3 1\right\rangle} & 
{\left\langle 2 1\right|H_N}{\left | 4 1\right\rangle} & \cdots
\\{\left\langle 3 1\right|H_N}{\left | 2 1\right\rangle} &
{\left\langle 3 1\right|H_N}{\left |3 1\right\rangle} &
{\left\langle 3 1\right|H_N}{\left | 4 1\right\rangle}& \cdots
\\{\left\langle 4 1\right|H_N}{\left | 2 1\right\rangle} &
{\left\langle 4 1\right|H_N}{\left | 3 1\right\rangle} &
{\left\langle 4 1\right|H_N}{\left | 4 1\right\rangle}& \cdots
 \\ \vdots & \vdots & \vdots & \ddots
  \end{pmatrix}
  =2
  \begin{pmatrix}
  1 & 0 & 0 & \cdots \\  0 & 1 & 0 & \cdots \\ 0 & 0 & 1 & \cdots
  \\ \vdots & \vdots & \vdots & \ddots
  \end{pmatrix}  
  \end{equation}
The matrix representation of the vertical ladder operators $A_\pm$  and $\rho$  is:
\begin{equation}
A_+=
  \begin{pmatrix}
  0 & 0 & 0 & 0 & \cdots \\  2 & 0 & 0 & 0 & \cdots 
  \\ 0 & \sqrt{10} & 0 & 0 & \cdots
  \\ 0 & 0 &  3\sqrt{2} & 0 & \cdots
  \\ \vdots & \vdots & \vdots & \vdots & \ddots
  \end{pmatrix},~
A_-=
  \begin{pmatrix}
  0 & 2 & 0 & 0 & \cdots \\  0 & 0 & \sqrt{10} & 0 & \cdots 
  \\ 0 & 0 & 0 & 3\sqrt{2} & \cdots
  \\ 0 & 0 & 0 & 0 & \cdots
  \\ \vdots & \vdots & \vdots & \vdots & \ddots
  \end{pmatrix},~
  \rho=
  \begin{pmatrix}
  4 & -2 & 0 & \cdots \\  -2 & 6 & -\sqrt{10} & \cdots 
  \\ 0 & -\sqrt{10} & 8 & \cdots
  \\ \vdots & \vdots & \vdots  & \ddots
  \end{pmatrix}
  \end{equation}
It is quite clear that the matrix representation of the operators 
$ P_\rho , \rho P_\rho $  is:
\begin{equation}
 P_\rho= \frac{i\hbar}{2}
    \begin{pmatrix} 
     -31/105 & 43/105 & 11\sqrt{10}/105 & \cdots  
  \\  -62/105 & -19/105 & 22\sqrt{10}/105 & \cdots 
  \\ -2\sqrt{10}/21 & -4\sqrt{10}/21 & 1/21 & \cdots
  \\ \vdots & \vdots  & \vdots & \ddots  
    \end{pmatrix},~~
\rho P_\rho= \frac{i\hbar}{2}
    \begin{pmatrix}
     0 & 2 & 0 & 0 & \cdots 
  \\  -2 & 0 & \sqrt{10} & 0 & \cdots 
  \\ 0 & -\sqrt{10} & 0 & 3\sqrt{2} & \cdots
  \\ 0 & 0 &  -3\sqrt{2} & 0 & \cdots
  \\ \vdots & \vdots & \vdots & \vdots & \ddots 
    \end{pmatrix}
\end{equation}
The joint eigenvectors of $N, H_N$ can be obtained according to Eq.~\ref{eq:28} as follows:
\begin{equation}
N \left \vert {n l} \right \rangle =
  \begin{pmatrix}
  2 & 0 & 0 & \cdots \\  0 & 3 & 0 & \cdots \\ 0 & 0 & 4 & \cdots
  \\ \vdots & \vdots & \vdots & \ddots
  \end{pmatrix}
  \begin{pmatrix}
  a \\  b \\ c \\ \vdots 
  \end{pmatrix}
  = n
  \begin{pmatrix}
 a \\  b \\ c \\ \vdots 
  \end{pmatrix}
  \Rightarrow \begin{cases}
    ~2a=n a \\ ~3b=n b \\ ~4c=n c \\ ~\vdots
  \end{cases} 
\end{equation}
The normalized solutions of these equations for $n=2,3,4,...\in \mathbb N$  are respectively given by:
 \begin{equation}
     \left \vert 2 1 \right \rangle =
     \begin{pmatrix}
      1 \\  0 \\ 0 \\ \vdots 
     \end{pmatrix},~
     \left \vert 3 1 \right \rangle =
     \begin{pmatrix}
      0 \\  1 \\ 0 \\ \vdots 
     \end{pmatrix},~
     \left \vert 4 1 \right \rangle =
     \begin{pmatrix}
      0 \\  0 \\ 1 \\ \vdots 
     \end{pmatrix}, ...
 \end{equation}
We can verify that these vectors are orthonormal, and according to Eq.~\ref{eq:22}, we have:
\begin{equation}
    \left \langle \acute{n}, 1 \vert n, 1 \right \rangle =
    \delta_{\acute{n}n},~
     \sum_{n=2}^N  
    \left \vert n,1 \right \rangle
    \left \langle n,1 \right \vert = 1
\end{equation}
\begin{table*}[ht!]
\section*{Appendix B:~Some arbitrary points of the first modified functions ${M_l}$}

\centering
\begin{small}

\begin{tabular}{c c c c c c c c|}
&\\

 $x$ & $M_0 $ & $M_1$ & $M_2$& $M_3$& $M_4$& $M_5$ \\

0.0 &~ 1.0000000000 &~ 2.4494897428 &~ 10.9544511501 &~ 70.9929573972 &~ 602.3952191045 &~ 6317.9743589223 \\

 0.1 &~ 0.9993753253 &~ 2.4487244447 &~ 10.9521692784 &~ 70.9818658067 &~ 602.3199252637 &~ 6317.3162761421\\

 0.2 &~ 0.9975051971 &~ 2.4464305576 &~ 10.9453273289 &~ 70.9486037406 &~ 602.0941104643 &~ 6315.3425022980\\

 0.3 &~ 0.9944012393 &~ 2.4426140912 &~ 10.9339362860 &~ 70.8932092849 &~ 601.7179747578 &~ 6312.0544602172\\

0.4 &~ 0.9900826177 &~ 2.4372850211 &~ 10.9180144144 &~ 70.8157458168 &~ 601.1918511720 &~ 6307.4545190678\\

 0.5 &~ 0.9845757344 &~ 2.4304572283 &~ 10.8975871963 &~ 70.7163018552 &~ 600.5162051248 &~ 6301.5459910489\\

0.6 &~ 0.9779138150 &~ 2.4221484155 &~ 10.8726872434 &~ 70.5949908532 &~ 599.6916336058 &~ 6294.3331267679\\

 0.7 &~ 0.9701363985 &~ 2.4123800015 &~ 10.8433541846 &~ 70.4519509323 &~ 598.7188641290 &~ 6285.8211093137\\

 0.8 &~ 0.9612887480 &~ 2.4011769941 &~ 10.8096345309 &~ 70.2873445593 &~ 597.5987534583 &~ 6276.0160470419\\

 0.9 &~ 0.9514211956 &~ 2.3885678432 &~ 10.7715815165 &~ 70.1013581684 &~ 596.3322861106 &~ 6264.9249650860\\

 1.0 &~ 0.9405884429 &~ 2.3745842767 &~ 10.7292549197 &~ 69.8942017288 &~ 594.9205726399 &~ 6252.5557956163\\

 1.1 &~ 0.9288488317 &~ 2.3592611192 &~ 10.6827208616 &~ 69.6661082602 &~ 593.3648477092 &~ 6238.9173668661\\

 1.2 &~ 0.9162636050 &~ 2.3426360964 &~ 10.6320515865 &~ 69.4173332973 &~ 591.6664679529 &~ 6224.0193909517\\

 1.3 &~ 0.9028961733 &~ 2.3247496277 &~ 10.5773252239 &~ 69.1481543072 &~ 589.8269096393 &~ 6207.8724505135\\

1.4 &~ 0.8888114010 &~ 2.3056446078 &~ 10.5186255334 &~ 68.8588700592 &~ 587.8477661365 &~ 6190.4879842073\\

 1.5 &~ 0.8740749261 &~ 2.2853661803 &~ 10.4560416350 &~ 68.5497999524 &~ 585.7307451919 &~ 6171.8782710784\\

 1.6 &~ 0.8587525237 &~ 2.2639615049 &~ 10.3896677245 &~ 68.2212833017 &~ 583.4776660309 &~ 6152.0564138539\\

1.7 &~ 0.8429095207 &~ 2.2414795214 &~ 10.3196027783 &~ 67.8736785851 &~ 581.0904562840 &~ 6131.0363211899\\

 1.8 &~ 0.8266102686 &~ 2.2179707104 &~ 10.2459502459 &~ 67.5073626562 &~ 578.5711487508 &~ 6108.8326889116\\

1.9 &~ 0.8099176768 &~ 2.1934868549 &~ 10.1688177348 &~ 67.1227299240 &~ 575.9218780094 &~ 6085.4609802887\\

 2.0 &~ 0.7928928100 &~ 2.1680808035 &~ 10.0883166874 &~ 66.7201915024 &~ 573.1448768809 &~ 6060.9374053879\\

 2.1 &~ 0.7755945465 &~ 2.1418062367 &~ 10.0045620520 &~ 66.3001743343 &~ 570.2424727586 &~ 6035.2788995481\\

2.2 &~ 0.7580792996 &~ 2.1147174388 &~ 09.9176719502 &~ 65.8631202913 &~ 567.2170838117 &~ 6008.5031010223\\

 2.3 &~ 0.7404007957 &~ 2.0868690762 &~ 09.8277673421 &~ 65.4094852536 &~ 564.0712150734 &~ 5980.6283278365\\

 2.4 &~ 0.7226099073 &~ 2.0583159835 &~ 09.7349716890 &~ 64.9397381725 &~ 560.8074544237 &~ 5951.6735539115\\

2.5 &~ 0.7047545368 &~ 2.0291129592 &~ 09.6394106188 &~ 64.4543601195 &~ 557.4284684786 &~ 5921.6583844998\\

2.6 &~ 0.6868795446 &~ 1.9993145702 &~ 09.5412115916 &~ 63.9538433234 &~ 553.9369983934 &~ 5890.6030309864\\

 2.7 &~ 0.6690267182 &~ 1.9689749685 &~ 09.4405035698 &~ 63.4386902007 &~ 550.3358555938 &~ 5858.5282851066\\
 
2.8 &~ 0.6512347755 &~ 1.9381477185 &~ 09.3374166922 &~ 62.9094123809 &~ 546.6279174439 &~ 5825.4554926318\\

2.9 &~ 0.6335393998 &~ 1.9068856365 &~ 09.2320819546 &~ 62.3665297296 &~ 542.8161228612 &~ 5791.4065265777\\

3.0 &~ 0.6159732983 &~ 1.8752406425 &~ 09.1246308968 &~ 61.8105693736 &~ 538.9034678908 &~ 5756.4037599850\\

 3.1 &~ 0.5985662837 &~ 1.8432636247 &~ 09.0151952982 &~ 61.2420647297 &~ 534.8930012477 &~ 5720.4700383298\\

 3.2 &~ 0.5813453711 &~ 1.8110043165 &~ 08.9039068825 &~ 60.6615545397 &~ 530.7878198385 &~ 5683.6286516137\\

3.3 &~ 0.5643348888 &~ 1.7785111855 &~ 08.7908970320 &~ 60.0695819156 &~ 526.5910642729 &~ 5645.9033061881\\

 3.4 &~ 0.5475565993 &~ 1.7458313364 &~ 08.6762965128 &~ 59.4666933962 &~ 522.3059143739 &~ 5607.3180963658\\

3.5 &~ 0.5310298264 &~ 1.7130104242 &~ 08.5602352117 &~ 58.8534380186 &~ 517.9355846977 &~ 5567.8974758725\\

 3.6 &~ 0.5147715879 &~ 1.6800925806 &~ 08.4428418844 &~ 58.2303664052 &~ 513.4833200724 &~ 5527.6662291893\\

 3.7 &~ 0.4987967293 &~ 1.6471203515 &~ 08.3242439171 &~ 57.5980298707 &~ 508.9523911648 &~ 5486.6494428387\\

 3.8 &~ 0.4831180593 &~ 1.6141346451 &~ 08.2045671003 &~ 56.9569795490 &~ 504.3460900844 &~ 5444.8724766648\\

 3.9 &~ 0.4677464831 &~ 1.5811746910 &~ 08.0839354160 &~ 56.3077655428 &~ 499.6677260332 &~ 5402.3609351557\\

 4.0 &~ 0.4526911341 &~ 1.5482780089 &~ 07.9624708387 &~ 55.6509360977 &~ 494.9206210098 &~ 5359.1406388593\\

 4.1 &~ 0.4379595018 &~ 1.5154803868 &~ 07.8402931497 &~ 54.9870368021 &~ 490.1081055758 &~ 5315.2375959378\\

 4.2 &~ 0.4235575563 &~ 1.4828158678 &~ 07.7175197649 &~ 54.3166098142 &~ 485.2335146919 &~ 5270.6779739097\\

 4.3 &~ 0.4094898668 &~ 1.4503167453 &~ 07.5942655769 &~ 53.6401931176 &~ 480.3001836315 &~ 5225.4880716215\\

 4.4 &~ 0.3957597168 &~ 1.4180135656 &~ 07.4706428102 &~ 52.9583198068 &~ 475.3114439775 &~ 5179.6942914951\\

 4.5 &~ 0.3823692120 &~ 1.3859351370 &~ 07.3467608903 &~ 52.2715174027 &~ 470.2706197104 &~ 5133.3231120923\\

 4.6 &~ 0.3693193842 &~ 1.3541085461 &~ 07.2227263254 &~ 51.5803072006 &~ 465.1810233922 &~ 5086.4010610367\\

 4.7 &~ 0.3566102881 &~ 1.3225591787 &~ 07.0986426025 &~ 50.8852036489 &~ 460.0459524526 &~ 5038.9546883320\\

 4.8 &~ 0.3442410937 &~ 1.2913107472 &~ 06.9746100948 &~ 50.1867137617 &~ 454.8686855820 &~ 4991.0105401148\\

 4.9 &~ 0.3322101723 &~ 1.2603853208 &~ 06.8507259825 &~ 49.4853365642 &~ 449.6524792366 &~ 4942.5951328764\\

 5.0 &~ 0.3205151771 &~ 1.2298033617 &~ 06.7270841858 &~ 48.7815625708 &~ 444.4005642600 &~ 4893.7349281876\\

\end{tabular}
\end{small}

\caption{\label{tab:table} The values called from the graphs of modified Bessel's functions-orders 0,1,2,3,4 and 5 of ${M_l}(x)= \left ( \frac{x}{2} 
\right )^ {l+1/2} \left [ I_{2l+1}(x) \right ] ^ {-1/2}$ are determined with extra precision of $10^{-10}$. While naturally because of singularity at $x=0$, these estimated values measure as a one-sided limit.}

\end{table*}
\newpage
\section*{Appendix C:~Computing the first radial wave functions 
$R_{n l}(r)$}
\indent

We will extract the wave functions related to the ground state $ \left \vert 1 0 \right \rangle $.    
Based on Eq.~\ref{eq:47}, and the normalization condition of Eq.~\ref{eq:21}, the annihilation operator $A_-$ can generate the following results:
\begin{equation}
    A_-\phi_{10}(\rho)=
    \left (-\rho \frac{d}{d \rho} -\frac{\rho}{2} \right ) \phi_{10}(\rho)=0 \
\end{equation}
Where $ \phi_{10}(\rho)= e^{-\rho/2} $, and since 
$ R_{10}(r)= 
\frac{1}{\sqrt{2}} \left ( \frac{2Z}{a_\circ} \right )^{3/2} 
\phi_{10}(\rho),~ \rho= \frac{2Z}{a_\circ} r $
then:
\begin{equation}
    R_{10}(r)= 
2 \left ( \frac{Z}{a_\circ} \right )^{3/2} e^{-\frac{Z}{a_\circ}r}
\end{equation}
 By acting the creation operator $ A_+ $ on 
 $ \phi_{10}(\rho) $, we can say that:
\begin{equation}
    A_+\phi_{10}(\rho)= 
    \left (\rho \frac{d}{d \rho} -\frac{\rho}{2}+2 \right ) \phi_{10}(\rho)= \sqrt{2} \phi_{20}(\rho)
\end{equation}
 Where $\phi_{20}(\rho)=\frac{1}{\sqrt{2}}(2-\rho)e^{-\rho/2}$  and since $ R_{20}(r)= 
\frac{1}{2} \left ( \frac{Z}{a_\circ} \right )^{3/2} 
\phi_{20}(\rho),~ \rho= \frac{Z}{a_\circ} r $   then:
\begin{equation}
    R_{20}(r)= 2 \left ( \frac{Z}{2a_\circ} \right )^{3/2}
    \left ( 1- \frac{Z}{2a_\circ}r \right )
e^{-\frac{Z}{2a_\circ}r}
\end{equation}
By acting the creation operator $ A_+ $  on $\phi_{20}(\rho),$  it is obvious that:
\begin{equation}
    A_+\phi_{20}(\rho)= 
    \left (\rho \frac{d}{d \rho} -\frac{\rho}{2}+3 \right ) \phi_{20}(\rho)= \sqrt{6} \phi_{30}(\rho)
\end{equation}
Where $\phi_{30}(\rho)=\frac{1}{\sqrt{3}}
\left(3-3\rho+\frac{\rho^2}{2}\right)e^{-\rho/2}$, and since
$ R_{30}(r)= 
\frac{1}{\sqrt{6}} \left ( \frac{2Z}{3a_\circ} \right )^{3/2} 
\phi_{30}(\rho),~ \rho= \frac{2Z}{3a_\circ} r $ then:
\begin{equation}
    R_{30}(r)= 2 \left ( \frac{Z}{3a_\circ} \right )^{3/2}
    \left [ 1- \frac{2Z}{3a_\circ}r+ \frac{2}{27} \left ( \frac{Z}{a_\circ} \right )^2 r^2 \right ]
     e^{-\frac{Z}{3a_\circ}r}
\end{equation}
Hence, the radial wave functions $ R_{10}(r),R_{20}(r),R_{30}(r) $ are obtained.

We will extract the radial wave functions related to the ground state $ \left \vert 2 1 \right \rangle $ that by the annihilation operator $ A_- $  and using the normalization condition, we have:
\begin{equation}
    A_-\phi_{2 1}(\rho)= 
    \left (-\rho \frac{d}{d \rho} -\frac{\rho}{2}+1 \right ) \phi_{21}(\rho)= 0
\end{equation}
Where  $\phi_{21}(\rho)=\frac{1}{\sqrt{6}} \rho e^{-\rho/2}$, and since $ R_{21}(r)= 
\frac{1}{2} \left ( \frac{Z}{a_\circ} \right )^{3/2} 
\phi_{21}(\rho),~ \rho= \frac{Z}{a_\circ} r $ then:
\begin{equation}
    R_{21}(r)= \frac{1}{\sqrt{3}}
    \left ( \frac{Z}{2a_\circ} \right )^{3/2}
    \left ( \frac{Z}{a_\circ}r \right ) e^{-\frac{Z}{2a_\circ}r}
\end{equation}
By acting the creation operator $ A_+ $  on $ \phi_{21}(\rho), $ the following result reads:
\begin{equation}
    A_+\phi_{21}(\rho)= 
    \left (\rho \frac{d}{d \rho} -\frac{\rho}{2}+3 \right ) \phi_{21}(\rho)= 2 \phi_{31}(\rho)
\end{equation}
Where $\phi_{31}(\rho)=\frac{1}{2\sqrt{6}} \rho (4-\rho)e^{-\rho/2}$, and since $ R_{31}(r)= \frac{1}{\sqrt{6}} 
\left ( \frac{2Z}{3a_\circ} \right )^{3/2} 
\phi_{31}(\rho),~ \rho= \frac{2Z}{3a_\circ} r $ then:
\begin{equation}
    R_{31}(r)= \frac{4\sqrt{2}}{9}
    \left ( \frac{Z}{3a_\circ} \right )^{3/2}
    \left ( \frac{Z}{a_\circ}r \right )
    \left (1- \frac{Z}{6a_\circ}r \right )e^{-\frac{Z}{3a_\circ}r}
\end{equation}
Hence, the radial wave functions $R_{21}(r), R_{31}(r)$ are obtained.

We can extract the radial wave function related to the ground state $ \left \vert 3 2 \right \rangle $ that using the annihilation operator $ A_- $ and the normalization condition is given by:
\begin{equation}
    A_-\phi_{32}(\rho)= 
    \left (-\rho \frac{d}{d \rho} -\frac{\rho}{2}+2 \right ) \phi_{32}(\rho)= 0
\end{equation}
Where $\phi_{32}(\rho)=\frac{1}{2\sqrt{30}} \rho^2 e^{-\rho/2}$, and since $ R_{32}(r)= \frac{1}{\sqrt{6}} 
\left ( \frac{2Z}{3a_\circ} \right )^{3/2} 
\phi_{32}(\rho),~ \rho= \frac{2Z}{3a_\circ} r $ then:
\begin{equation}
    R_{32}(r)= \frac{2\sqrt{2}}{27\sqrt{5}}
    \left ( \frac{Z}{3a_\circ} \right )^{3/2}
    \left ( \frac{Z}{a_\circ}r \right )^2 e^{-\frac{Z}{3a_\circ}r}
\end{equation}
Hence, the radial wave function $ R_{32}(r) $ is obtained.
\end{widetext}


\providecommand{\noopsort}[1]{}\providecommand{\singleletter}[1]{#1}%

\end{document}